\documentclass[12pt]{elsarticle}
\usepackage{graphicx}% Include figure files
\usepackage{dcolumn}% Align table columns on decimal point
\usepackage{bm}% bold math
\usepackage{url}
\usepackage{makecell}
\usepackage{amssymb}
 \expandafter\let\csname equation*\endcsname\relax 
  \expandafter\let\csname endequation*\endcsname\relax 
\usepackage{amsmath}
\usepackage[colorlinks=true, a4paper=true, pdfstartview=FitV,
linkcolor=blue, citecolor=blue, urlcolor=blue]{hyperref}

\include{epsf}

\journal{Renewable Energy}

\begin{document}

\begin{frontmatter}

\title{Wind speed forecasting at different time scales: a non parametric approach}
\author{Guglielmo D'Amico}  
\address{Dipartimento di Farmacia, 
Universit\`a `G. D'Annunzio' di Chieti-Pescara,  66013 Chieti, Italy}
\author{Filippo Petroni}
\address{Dipartimento di Scienze Economiche ed Aziendali,
Universit\`a degli studi di Cagliari, 09123 Cagliari, Italy}
\author{Flavio Prattico}
\address{Dipartimento di Ingegneria Industriale e dell'Informazione e di Economia, Universit\`a degli studi dell'Aquila, 67100 L'Aquila, Italy}
\bigskip

\begin{abstract}
The prediction of wind speed is one of the most important aspects when dealing with renewable energy. In this paper we show a new nonparametric model, based on semi-Markov chains, to predict wind speed. Particularly we use an indexed semi-Markov model, that reproduces accurately the statistical behavior of wind speed, to forecast wind speed one step ahead for different time scales and for very long time horizon maintaining the goodness of prediction. In order to check the main features of the model we show, as indicator of goodness, the root mean square error between real data and predicted ones and we compare our forecasting results with those of a persistence model.
\end{abstract}


\date{\today}

\end{frontmatter}

\section{Introduction}
The variations of wind speed, in a certain site, are strictly related to the economic aspects of a wind farm, such as maintenance operations, especially in the off shore farms, pitch angle control on new wind turbines and evaluation of a new site. Many researchers are working proposing new models that can allow the prediction of wind speed, minutes, hours or days ahead. Many of these models are based on neural networks \cite{14,03}, autoregressive models \cite{05,07,753}, Markov chains \cite{sha05,nfa04,06,01,13,794}, hybrid models where the previous mentioned models are combined \cite{you03,01,08,10,11,16,480,555,fore4,fore6,fore8,fore13,f1,f2} and other less used models \cite{fore1,fore2,fore3,fore7,fore9,fore11}. Often, these models are either focused on specific time scale forecasting, or synthetic time series generation. Instead, our model can be used both for time series generation and for forecasting at different time scales.

The approach we propose here is based on indexed semi-Markov chain (ISMC) model that was advanced by the same authors in \cite{wind2} and applied to the generation of synthetic wind speed time series. In \cite{wind2} we showed that our model is able to reproduce correctly the statistical behavior of wind speed. The ISMC model is a nonparametric model because it does not require any assumption on the form of the distribution function of wind speed.
In this work we use the same model, slightly modified by adding a daily deterministic component, to forecast future values of wind speed. We will show that this model performs better than a simple persistence model, by comparing the root mean square errors. The  ISMC model is able to forecast wind speed at different time scale without loosing the goodness of forecasting which is almost independent from the time horizon. Another important aspect addressed by this work is the number of data needed to have a good forecast. With this aim we will show the root mean square error as a function of the data used to calibrate the model.  

The paper is organized as follows. First of all, in Section 2, we describe the database used for the analysis. In Section 3, we present the model and its validation. Then, in Section 4, we present results of the wind speed forecasting through an indicator of goodness and comparison with the persistence model. Finally in Section 5 we present some concluding remarks.

\section{Database}
\begin{figure}
\centering
\includegraphics[height=10cm]{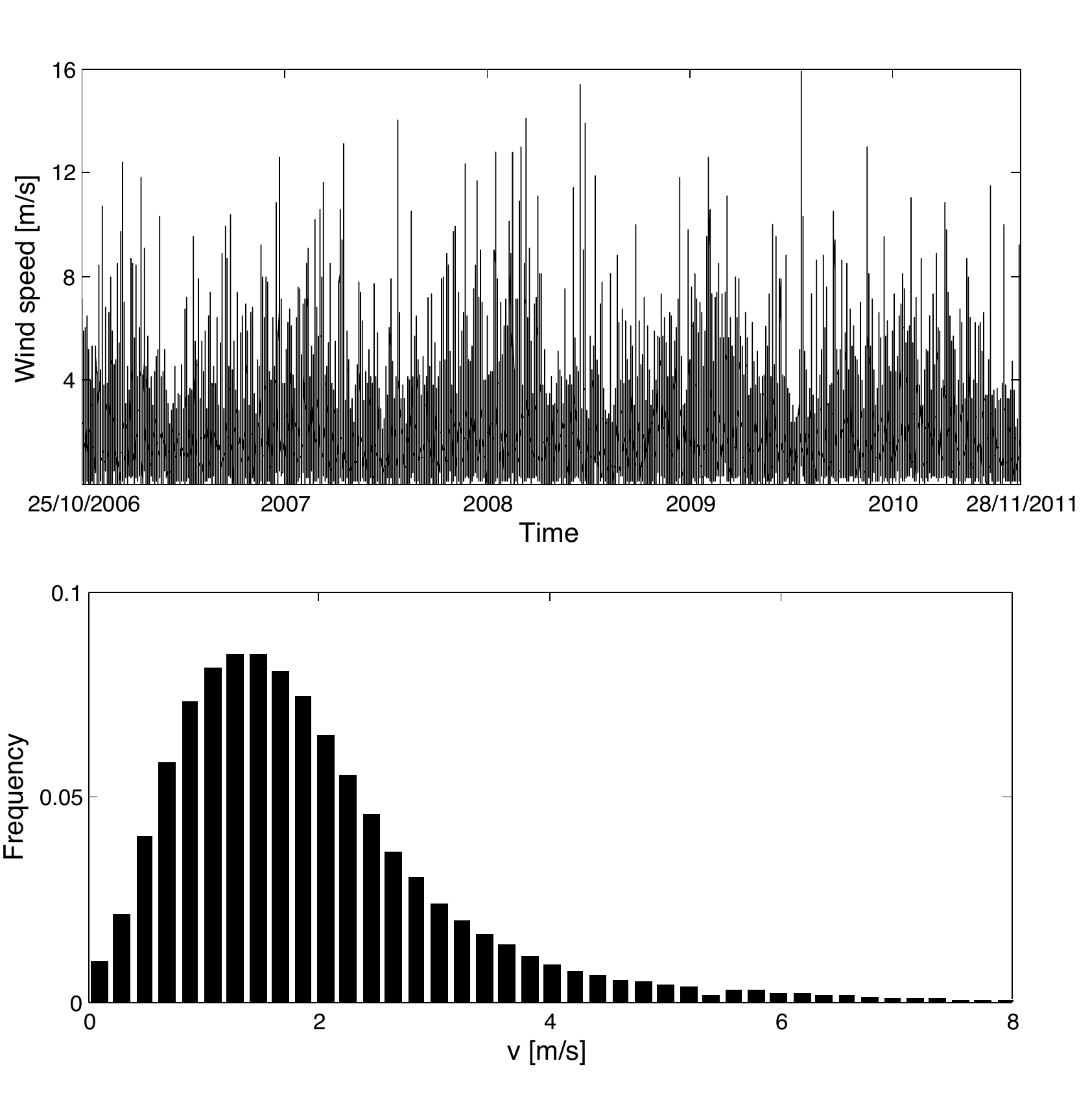}
\caption{Database and its probability density distribution.}\label{fig1}
\end{figure}

The database used for the analysis in this work is freely available from \cite{data} and is composed of more than 230000 data of wind speed collected every 10 minutes. The weather station of L.S.I. -Lastem is situated in Italy at N 45$°$ 28' 14,9'' $-$ E 9$°$ 22' 19,9'' and at 107 $m$ of altitude. The station uses a combined speed-direction anemometer at 22 $m$ above the ground. It has a measurement range that goes from 0 to 60 $m/s$, a threshold of 0,38 $m/s$ and a resolution of 0,05 $m/s$. The database and its empirical probability density function are represented in Figure \ref{fig1}. 
We discretized wind speed into 8 states (see Table \ref{st}) chosen to cover all the wind speed distribution. Table \ref{st} shows the wind speed states with their related wind speed values.

\begin{table}
\begin{center}
\begin{tabular}{|c|*{2}{c|}|}
     \hline
Sate & Wind speed range $m/s$  \\ \hline
1 & 0 to 1  \\ \hline
2 & 1 - 2  \\ \hline
3 & 2 - 3  \\ \hline
4 & 3 - 4 \\ \hline
5 & 4 - 5  \\ \hline
6 & 5 - 6 \\ \hline
7 & 6 - 7  \\ \hline
8 & $>$7  \\ \hline
\end{tabular} 
\caption{Wind speed discretization}
\label{st} 
\end{center}
\end{table}

In order to analyze the behavior at different time scales, we resampled the data at different sampling frequencies: namely 30 minutes, 1 hour and 2 hours. 

\section{Model}

\subsection{The indexed semi-Markov chain model}

The general formulation of the ISMC as developed in references \cite{dami11a}, \cite{dami11b}, \cite{dami12b} and \cite{wind2} is here discussed informally.

Semi-Markov processes have similar idea as those that generate Markov processes. The processes are both described by a set of finite states $v_n$ whose transitions are ruled by a transition probability matrix. The semi-Markov process differs from the Markov process because the transition times $T_n$ are generated according to random variables. Indeed, the time between transitions $T_{n+1}-T_n$ is random and may be modeled by means of any type of distribution functions.
In studies concerning wind speed modeling the states $v_n$ indicates discretized wind speed at the nth transition and $T_n$ the time in which the nth change of wind speed occurs. 

In \cite{dami12,wind4}, different semi-Markov models were applied to the wind speed modeling and it was shown that the semi-Markov models over perform the Markov models and therefore they are to be preferred in the modeling of wind speed to Markovian models. 

In order to better represent the statistical characteristics of wind speed, in a recent article, the idea of an ISMC was advanced in the field of wind speed, see \cite{wind2}. 
The novelty, with respect to the semi-Markov case, consists in the introduction of a third random variable  defined as follow:
\begin{equation}
U_{n}^{m}= \sum_{k=0}^{m} v_{n-1-k} \cdot \frac{T_{n-k}-T_{n-1-k}}{T_{n}-T_{n-1-m}}. 
\end{equation}
This variable can be interpreted  as a moving average of order $m+1$ executed on the series of the past wind speed values $(v_{n-1-k})$ with weights given by the fractions of sojourn times in that wind speed $(T_{n-k}-T_{n-1-k})$ with respect to the interval time on which the average is executed $(T_n-T_{n-1-m})$.
Also the process $U^m$ has been discretized, Table \ref{Um} shows the states of the process and their values.

\begin{table}
\begin{center}
\begin{tabular}{|c|*{2}{c|}|}
     \hline
Sate & $U^m$ range $m/s$  \\ \hline
1 & 0 to 2.1  \\ \hline
2 & 2.1 - 2.6  \\ \hline
3 & 2.6 - 3.4  \\ \hline
4 & 3.4 - 6 \\ \hline
5 & $>$6  \\ \hline
\end{tabular} 
\caption{$U^m$ processes discretization}
\label{Um} 
\end{center}
\end{table}

The parameter $m$ must be optimized as a function of the specific database. The optimization is made by finding the value of $m$ that realize the minimum of the root mean square error (RMSE) between the autocorrelation functions (ACF) of real and simulated data, see \cite{wind2}. In our analysis $m=7$.

The reasons to introduce this index of memory are found in the presence of a strong autocorrelation that characterize the wind speed process.
In the same work we have shown that if a too small memory is used, the autocorrelation is already persistent but decreases faster than real data. With a longer memory the autocorrelation remain high for a very long period and also its value is very close to that of real data. If $m$ is increased further the autocorrelation drops again to small values. This behavior suggests the existence of an optimal memory $m$. In our opinion one can justify this behavior by saying that short memories are not enough to identify in which status (low, medium low, medium, medium high, high, see Table 2) is the index $U^m$, too long memories mix together different status and then much of the information is lost in the average.

The one step transition probability matrix can be evaluated by considering the counting transition between the three random variables considered before. Then, the probability $p_{i,j}(t,u)$ represents the transition probability from the actual wind speed state $i$, to the wind speed state $j$, given that the sojourn time spent in the state $i$ is equal to $t$ and the value of the process $U^m$ is $u$. These probabilities can be computed as:

\begin{equation}
p_{i,j}(t,u)= \frac{ n_{i,j} (t,u) }{\sum\limits_{j} n_{i,j}(t,u)},
\end{equation}
\label{pri}

\noindent where $n_{ij}(t,u)$ is the total number of transitions observed in the database from state $i$ to state $j$ in next period having a sojourn time spent in the wind speed $i$ equal to $t$ and the value of the index process  equal to $u$.

\indent The ISMC model revealed to be particularly efficient in reproducing together the probability density function of wind speed and the autocorrelation function, see \cite{wind2}.
 
\subsection{Deterministic wind speed component}
\label{deter}
The speed of wind shows a diurnal behavior due to the alternation between night and day. In Figure \ref{acfp}, in which are plotted the ACF of real and simulated data, it is possible to note this sinusoidal trend (see Section \ref{nodo} for a better explanation of the figure). To model this seasonality we add a deterministic component given by a sine wave to the indexed semi-Markov model: 
\begin{equation}
v_d = A \cdot sin \left( \frac{2 \pi}{24} h \right) \;\;\;\; h=1,2,...,n.
\end{equation}
\label{det}
The value of the parameter $A$ has to be optimized according to the database used for the analysis. In our case $A=0.41$ and it has been obtained by minimizing the RMSE between the ACF of real and synthetic data by using a genetic algorithm.

\subsection{Transition probability matrix}
We computed the transition probability matrix by using equation $(2)$ to the wind speed database. Two examples of the estimated matrices  are given in Tables \ref{m1} and \ref{m2}. As described above, in the model the transition matrix do depend from initial and arrival states but also from the sojourn time and the value of the random variable U. In the example given here we show the transition matrices for $U^m=2$ and $t=2$ and for $U^m=4$ and $t=2$ respectively, evaluated from the original database with the sampling frequency of 10 minutes.

A first comparison between Table \ref{m1} and Table \ref{m2} reveals that the value of the index process  affects seriously the transition probability to the next wind speed value. As a matter of example if $i=1$, $t=2$, $U_{n}^{m}=2$, the probability to have a wind speed $j=1$ in next period is equal to $0.7065$, see Table \ref{m1}. On the contrary, if $i=1$, $t=2$, $U_{n}^{m}=4$, the probability to have a wind speed $j=1$ in next period becomes $0.4900$, see Table \ref{m2}. The differences in the one step transition probabilities are significant and confirm the hypothesis that next wind speed depends also on the value of the index process. This fact shows that the index process should be used when dealing with wind speed data.

\begin{table}
\begin{center}
\centering

\begin{tabular}{|l|c|c|c|c|c|c|c|c|}
\hline
$P_{ij}$&\textbf{j=1}&\textbf{2}&\textbf{3}&\textbf{4}&\textbf{5}&\textbf{6}&\textbf{7}&\textbf{8}\\\hline
\textbf{i=1}&0.7065&0.2856&0.0074&0.0001&0.0001&0.0001&0.0001&0.0000\\\hline
\textbf{2}&0.1546&0.7095&0.1310&0.0042&0.0004&0.0002&0.0001&0.0001\\\hline
\textbf{3}&0.0064&0.2779&0.6300&0.0800&0.0045&0.0008&0.0003&0.0003\\\hline
\textbf{4}&0.0005&0.0170&0.3227&0.5764&0.0773&0.0044&0.0011&0.0005\\\hline
\textbf{5}&0.0000&0.0054&0.0349&0.3737&0.4919&0.0753&0.0134&0.0054\\\hline
\textbf{6}&0.0000&0.0000&0.0000&0.0238&0.4048&0.3929&0.1786&0.0000\\\hline
\textbf{7}&0.0000&0.0000&0.0357&0.0357&0.0714&0.3571&0.2857&0.2143\\\hline
\textbf{8}&0.0000&0.0000&0.0000&0.0000&0.0435&0.0000&0.2174&0.7391\\\hline
\end{tabular}
\caption{Transition matrix for $U^m=2$ and $t=2$.}\label{m1}
\end{center}
\end{table}

\begin{table}
\begin{center}
\centering

\begin{tabular}{|l|c|c|c|c|c|c|c|c|}
\hline
$P_{ij}$&\textbf{j=1}&\textbf{2}&\textbf{3}&\textbf{4}&\textbf{5}&\textbf{6}&\textbf{7}&\textbf{8}\\\hline
\textbf{i=1}&0.4900&0.4300&0.0700&0.0000&0.0100&0.0000&0.0000&0.0000\\\hline
\textbf{2}&0.1002&0.6171&0.2488&0.0242&0.0048&0.0016&0.0000&0.0032\\\hline
\textbf{3}&0.0048&0.1456&0.6323&0.1975&0.0185&0.0000&0.0014&0.0000\\\hline
\textbf{4}&0.0000&0.0154&0.2270&0.5886&0.1553&0.0113&0.0018&0.0006\\\hline
\textbf{5}&0.0000&0.0009&0.0268&0.2763&0.5638&0.1220&0.0092&0.0009\\\hline
\textbf{6}&0.0000&0.0000&0.0060&0.0301&0.3414&0.5120&0.1004&0.0100\\\hline
\textbf{7}&0.0000&0.0000&0.0000&0.0000&0.0467&0.3400&0.4600&0.1533\\\hline
\textbf{8}&0.0000&0.0000&0.0116&0.0233&0.0000&0.0465&0.2674&0.6512\\\hline
\end{tabular}
\caption{Transition matrix for $U^m=4$ and $t=2$.}\label{m2}
\end{center}
\end{table}

\subsection{Model validation}
\label{nodo}
\begin{figure}
\centering
\includegraphics[height=8cm]{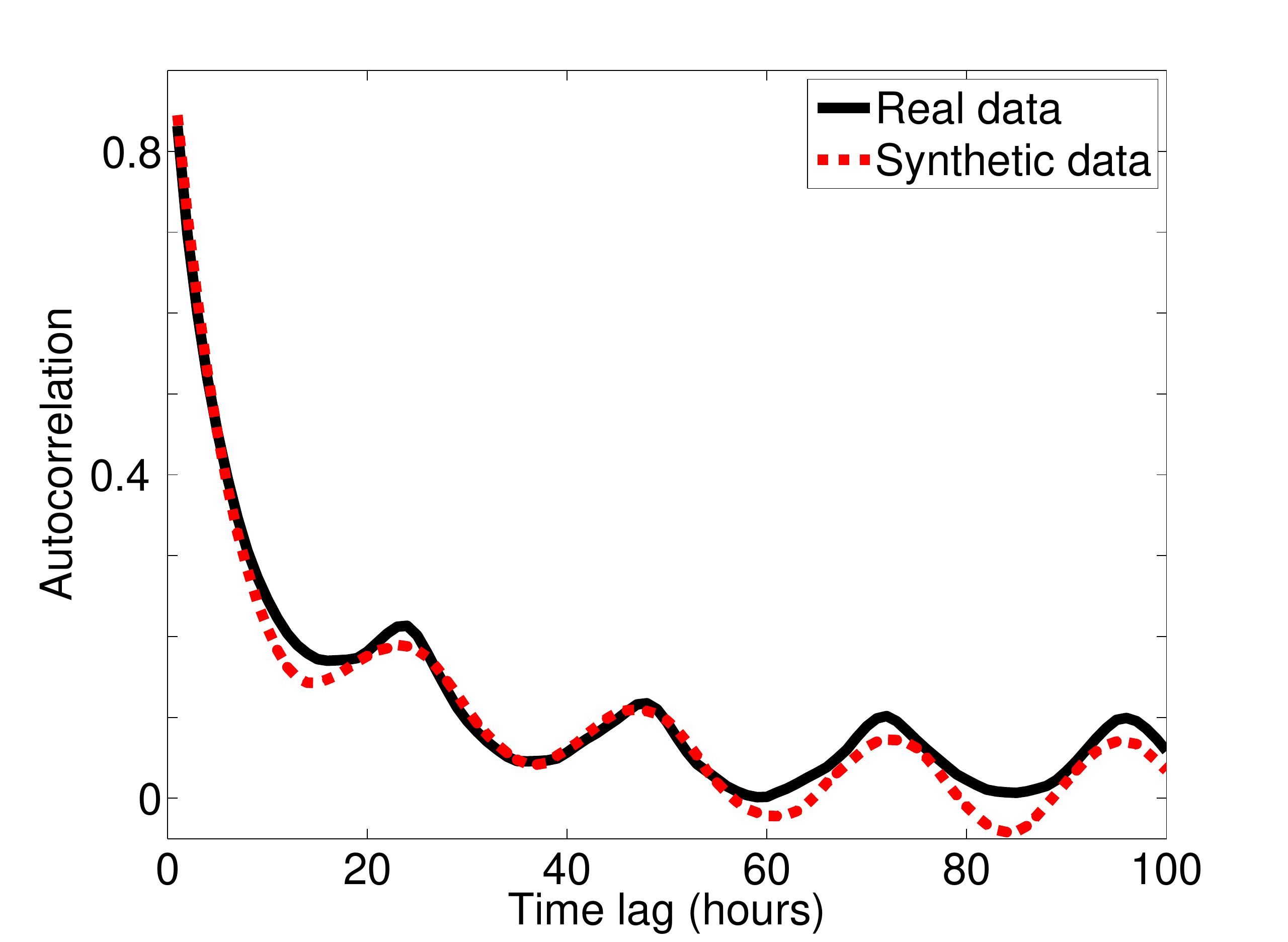}
\caption{Autocorrelation function of real and synthetic data}\label{acfp}
\end{figure}

We compute the ACF of real and synthetic data in order to assess the ability of the model to reproduce statistical properties of real wind speed data. We generate a synthetic time series by means of Monte Carlo simulation. The specific algorithm used for the generation of the trajectory can be found in \cite{wind2}.
If $v$ indicates wind speed, the time lagged $(\tau)$ autocorrelation of wind speed is defined as:
\begin{equation}
\label{autosquare}
\Sigma(\tau)=\frac{Cov(v(t+\tau),v(t))}{Var(v(t))}.
\end{equation}
The time lag $\tau$ was made to run from 10 minutes up to 100 hours. 
The ACF of real and the synthetic data are plotted in Figure \ref{acfp}.  As it is possible to note, the ACF has a sinusoidal trend with a period of 24 hours. This behavior is reproduced by our model with the introduction of the deterministic wind speed component evaluated by the equation $(\ref{det})$.
To asses the differences between the ACF of real and synthetic data we used the root mean square error (RMSE) which is defined as follows:
$$ RMSE =  \sqrt{ \frac{1}{n} \sum_{i=1}^n \left( v_{i}^{r} - v_{i}^{s} \right)^2 } , $$
where $v^{r}$ and $v^{s}$ represent real data and synthetic one respectively, while $n$ is the length of the two series. For the ACF plotted in Figure \ref{acfp} we obtained a RMSE equal to $0.0223$.

\section{Results}
\subsection{Wind speed forecasting}
In this section the ISMC model is used to forecast future wind speed states by using a one step ahead forecasting procedure,  for different time horizons and for various time scales. Particularly, we tested our model using the previously described databases with a sampling frequency of 10 minutes, 30 minutes, 1 hour and 2 hours. 

For each one of the sampling frequencies, the database is divided into two subsets: the first part is used to find the transition probability matrix (as described in the previous section), we will call this part the setting period; the second part is used to compare the model forecasting with real data (called testing period).
As a first attempt to verify the model performance, we used two years of data as setting period and one year as testing. We will show in the paragraph \ref{nod} how to find the best setting period.
Once the transition matrix is set, the forecasted states are computed as follows:
\begin{equation}
v^f=\sum_{j=1}^{k} j \, p_{i,j}(t,u),
\end{equation}
where $k$ is the number of states in which wind speed is discretized and $p_{i,j}(t,u)$ is the transition probability matrix. The formula represents the expected value of the next transition given that the present wind speed value is $i$, the sojourn time spent in the state $i$ is equal to $t$ and the value of the index process $U^{m}$ is $u$. 

\begin{figure}
\centering
\includegraphics[height=10cm]{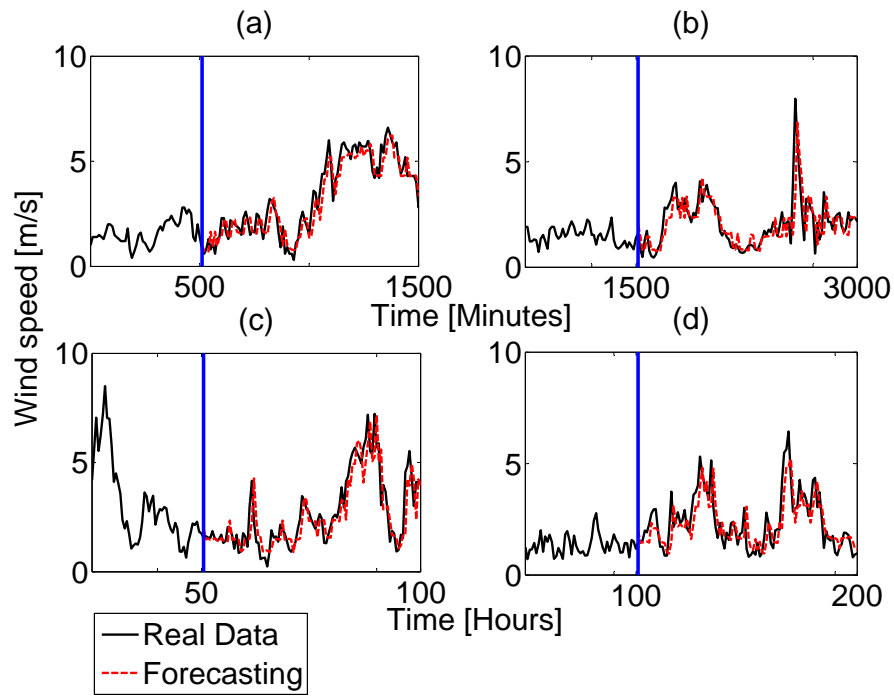}
\caption{Wind speed forecasting one step ahead for 100 time horizon. (a) 10 minutes database, (b) 30 minutes database, (c) 1 hour database, (d) 2 hours database.}\label{figone}
\end{figure}

In Figure \ref{figone} we show the results obtained using our model for the four different time scales. In the figure the black continuous line represents real data while the dashed red line is the predicted series. In this figure the predicted series are long 100 time horizon (specific time depending on the sampling frequency). 

Already from this figure, it is possible to note that the goodness of the prediction does not fall down at the increasing of the length of the forecasted series. 
To better verify this point, in Table \ref{tabE} we show quantitative results of our forecasting model for all the considered time scales and for different time horizons. Particularly, we show mean and standard deviation of the RMSE between real and predicted data tested on 50 different forecasted series. Table \ref{tabE} shows that the goodness of prediction remains almost constant even varying time scales and time horizons.
\begin{table}
\begin{center}
\begin{tabular}{|l|c|c|c|c|}\hline
\diaghead{\theadfont Diag ColumnmnHead II}%
{Time\\Scale}{Time\\Horizon}&
\thead{50}&\thead{100}&\thead{500}& \thead{1000}\\
\hline
10 minutes & 0.44 $\pm $ 0.02 & 0.44 $\pm $ 0.02 & 0.48 $\pm $ 0.02 & 0.52 $\pm $ 0.02 \\
\hline
30 minutes & 0.48 $\pm $ 0.01  & 0.50 $\pm $ 0.01  & 0.56 $\pm $ 0.01 & 0.62 $\pm $ 0.01 \\
\hline
1 hour & 0.54 $\pm $ 0.01 & 0.54 $\pm $ 0.01 & 0.61 $\pm $ 0.01 & 0.64 $\pm $ 0.01 \\
\hline
2 hour & 0.56 $\pm $ 0.01 & 0.59 $\pm $ 0.01 & 0.65 $\pm $ 0.01 & 0.69 $\pm $ 0.01 \\
\hline
\end{tabular}
\end{center}
\caption{RMSE between real data and forecasted series for different time scale and time horizon.}
\label{tabE}
\end{table}

We compare our model with a simple persistence model. This simple method is often used, still today, in industry for its simplicity and for its efficiency for very short-term predictions. It assumes that the wind speed at time $t+ \Delta t$ is equal to the wind speed at time $t$. Commonly this method is used to compare the behavior of new forecasting models \cite{pers}. Overall our model has a higher efficiency in the forecast for all the time scales and time horizons. The persistence model do not change its goodness of forecasting at varying of the time horizon. Then we compare our results with the persistence model at different time scales. For the frequency of 10 minutes, 30 minutes, 1 hour and 2 hours we have respectively an RMSE between the true series and the forecasted one generated through the persistence model of $0.59 \pm 0.05$, $0.63 \pm 0.08$, $0.73 \pm 0.09$ and $0.85 \pm 0.11$. As is possible to note the persistence model has less precision on the forecasting of the wind speed with respect to our model and the standard deviation increases at the increasing of the time scale in contrast to our model that has a reduction of the variability at the increasing of the time scale.

\subsection{Number of data optimization}\label{nod}
A serious problem to deal with in applying a nonparametric model is that of data availability. An important point is that of establishing the dimension of the setting period needed for a correct implementation of the model. From one part, reducing the setting period may determine the goodness of prediction to drop down; on the other hand the availability of large database is time consuming and consequently not economically efficient and sometimes not statistically necessary. To fix this point as related to the ISMC model we computed the RMSE between real data and a forecasted time series of 1000 time horizon. 

\begin{figure}
\centering
\includegraphics[height=8cm]{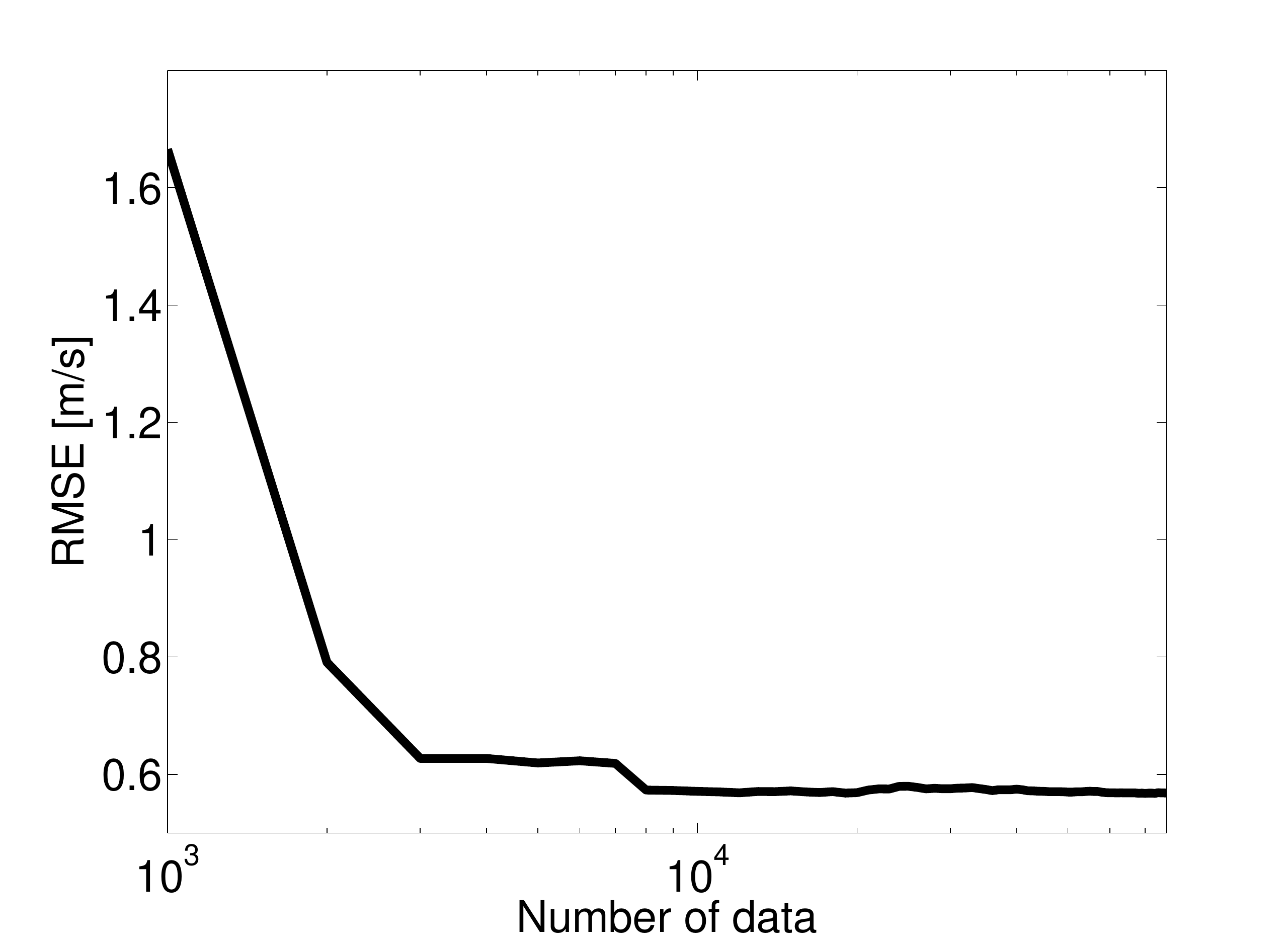}
\caption{RMSE between real wind speed and forecasted series as a function of the logarithm of the number of data.}\label{mincc}
\end{figure}

We show, in Figure \ref{mincc}, the results obtained for the 30 minutes sampling frequency. It can be noted that the RMSE, plotted as a function of the logarithm of the setting period length, after about 3000 data (corresponding roughly to 2 months) remains almost constant,  suggesting that the use of a larger setting period is not necessary. 
We repeated the same analysis for all the sampling frequency used obtaining: 20000 data (roughly 6 months) for 10 minutes sampling frequency, 2500 data (roughly 3 months) for 1 hour, and 2000 (roughly 5 months) for 2 hours. The decreasing in the number of data need to have a good forecasting is mainly due to the reduction of noise when the sampling frequency increases.

\section{Discussion and conclusion}
In previous works we presented new stochastic models, all based on a semi-Markov approach, to generate synthetic time series of wind speed. 
We showed that all the models perform better than corresponding Markov chain based models in reproducing statistical features of wind speed. Using these results, here, we tried to apply the model which we recognized to be the best among those, namely the indexed semi-Markov chain (ISMC) model, to forecast future wind speed in a specific site. 
The ISMC model is a nonparametric model and because of this it does not need any assumption on the distribution of wind speed and on wind speed variations.

In previous papers we showed that the ISMC model is able to reproduce correctly, and at the same time, both the probability distribution function of wind speed and the autocorrelation function. 

The results presented in this paper show that the model can be efficiently used to forecast wind speed at different horizon times. The forecast performance is almost independent from the time horizon used to forecast; the model can be used without degradation during the considered horizon time, at different time scales (we showed this for time scales ranging from 10 minutes to 2 hours).

The number of data needed to reach a good forecast performance do depend on the time scale used for forecasting; the model always works better than a simple persistence model.

All these characteristics suggest that the advanced ISMC model may be used both for modeling wind speed data and for wind speed prediction. Therefore, it may be utilized as input data for any wind energy system.

\end{document}